\documentstyle[twoside,fleqn,espcrc2]{article}

\def\be{\begin{equation}}
\def\ee{\end{equation}}
\def\bea{\begin{eqnarray}}
\def\eea{\end{eqnarray}}
\def\bed{\begin{displaymath}}
\def\eed{\end{displaymath}}
\def\nn{\nonumber}
\def\half{\frac{1}{2}}
\def\a{\alpha}
\def\b{\beta}

\def\d{\partial}
\def\de{\delta}
\def\e{\epsilon}           
\def\g{\gamma}

\def\l{\lambda}
\def\m{\mu}
\def\n{\nu}
\def\o{\omega}
\def\r{\rho}                      
\def\s{\sigma}                    
\def\t{\tau}

\def\G{\Gamma}

\def\S{\Sigma}

\def\calL{{\cal L}}

\def\R{{\rm R}}

\def\widebar{\overline}
\def\ul{\underline}
\def\ol{\widebar}

\def\psibar{\widebar{\psi}}
\def\phibar{\widebar{\phi}}

\def\one{{\bf 1}}

\def\dehat{\what{\de}}

\newcommand{\wt}[1]{\widetilde{#1}}
\newcommand{\what}[1]{\widehat{#1}}

\newcommand{\AmS}{{\protect\the\textfont2
  A\kern-.1667em\lower.5ex\hbox{M}\kern-.125emS}}

\title{\vspace{-15mm}
{\normalsize\null\hfill HUB-EP 97/11\\
\vspace{-4mm}
\null\hfill ITP-SB-97-15\\
\vspace{-4mm}
\null\hfill hep-th/9702123\\
\vspace{-4mm}
}
       $Osp(1|8)$-Gravity
       \thanks{To
       appear in the Proceedings of the 30th Ahrenshoop Symposium
       on the Theory of Elementary Particles, Nuclear Physics B,
       Proceedings Supplement, edited by D. L\"ust, 
       H.-J. Otto and G. Weigt. 
       Research supported by NSF grant Phy 9309888
       and the Swiss National Science Foundation.}}
       
\author{C.R. Preitschopf, 
  \address{Institut f\"ur Physik,\\
    Humboldt-Universit\"at zu Berlin,\\
    Invalidenstr. 110, D-10115 Berlin, Germany}
  T. Hurth${}^{{\rm b}}$, 
  P. van Nieuwenhuizen${}^{{\rm b}}$ 
  and A. Waldron
  \address{Institute for Theoretical
    Physics, State University of New York at Stony Brook\\ 
    Stony Brook, NY 11794-3840, USA}}%

\begin{document}
\newcommand{\rf}[1]{~(\ref{#1})}
\setlength{\topsep}{0.5 \topsep}
\setlength{\textfloatsep}{0.5 \textfloatsep}
\setlength{\intextsep}{0.5 \intextsep}

\thispagestyle{empty}

\begin{abstract}
We analyze a new MacDowell-Mansouri $R^2$-type supergravity
action based on the superalgebra $Osp(1|8)$. \hfill\break\indent
This contribution summarizes the work of ref.~\cite{ours}.

\end{abstract}

\maketitle

Recently, there has been renewed speculation that further supergravity
theories might exist in $d=11$ and $d=12$
dimensions~\cite{azcar}-\cite{11and12}, which might provide a unifying
origin for the supergravities appearing as low-energy limits of
string- or M-theory. The idea of supergravities beyond d=11 was
already explored fifteen years ago, but no {\it conventional}
supergravity theory was found~\cite{unconventional}, even though in
$d=(10,2)$ dimensions Majorana-Weyl spinors exist, and dimensional
reduction to $d=(10,1)$ would therefore lead to a $N=1$ supergravity
theory. The search for principles underlying string theory leads us to
investigate the superalgebraic foundations of supergravities. While
these are well understood in simple and usually low-dimensional cases,
for higher N or higher dimension one only suspects that they exist.
The indications we have of relations between all superstrings prompt
us to reopen the investigation.


\section{Algebras}

The recent spate of such ideas is based on the fact that p-branes
couple naturally to (p+1)-form gauge potentials via the
currents~\cite{azcar,pktrev}:
\be
\begin{array}[t]{l}
J^{\m_1 \ldots \m_{p+1}}(x) = 
\frac{1}{\sqrt{g}} \int d\t \int d^p \s \\
\phantom{mmm}\de^d (x - X(\t,\s) ) \ \e^{i_1 \ldots i_{p+1}}
\end{array}
\ee
\bed
\begin{array}[t]{l}
\phantom{mmm}\d_{i_1} X^{\m_1}(\t, \s) 
       \ldots  \d_{i_{(p+1)}} X^{\m_{(p+1)}}(\t, \s) \ .
\end{array}
\eed
\vspace{-1mm}
These currents are conserved:
\be
\d_\n ( \sqrt{g} J^{\n \m_1 \ldots \m_{p}}(x) ) \ = \ 0
\ee
and give rise to tensor charges
\be
Z^{\m_1 \ldots \m_{p}} \ = \ \int d^{d-1} x J^{0 \m_1 
\ldots \m_{p}}(x) \ .
\ee
They appear then in (maximally) extended supersymmetry algebras
as follows~\cite{pktdem}: for the case of IIA supersymmetry 
in d=(1,9) one has
\be
\begin{array}[t]{rl}
\{ Q_{\a} , Q_{\b}\} =& 
\G_{\a\b}^\m P_\m + \G_{\a\b}^\m Z_\m \\[2mm]
&+ 
\G_{\a\b}^{\m_1 \ldots \m_5} Z^+_{\m_1 \ldots \m_5} 
\\[2mm]
\{ Q^{\a} , Q^{\b}\} =&
\G_\m^{\a\b} P^\m - \G_\m^{\a\b} Z^\m \\[2mm]
&+ 
\G^{\m_1 \ldots \m_5 \a\b} Z^-_{\m_1 \ldots \m_5} 
\\[2mm]
\{ Q_{\a} , Q^{\b}\} =& 
\de_\a^\b Z + \G^{\m\n}{}_\a{}^\b Z_{\m\n} \\[2mm]
&+ 
\G^{\m_1 \ldots \m_4}{}_\a{}^\b Z_{\m_1 \ldots \m_4} 
\end{array} 
\label{twoa}
\ee
where the 16-dimensional Majorana-Weyl spinors $Q_a$ and
$Q^a$ have opposite chirality.  
This algebra may be given a (1,10)-d 
interpretation in terms of real
32-component spinors:
\bea
\{ Q_{a} , Q_{b}\} &=&
\begin{array}[t]{l}
\G_{ab}^M P_M + \G_{ab}^{M N} Z_{MN}\\[2mm] 
+ \G_{ab}^{M_1 \ldots M_5} Z_{M_1 \ldots M_5}\ , 
\end{array}
\label{elevenalg}
\eea
or even a (2,10)-d one in terms of 32-component Majorana-Weyl
spinors:
\bea
\{ Q_{a} , Q_{b}\} &=&
\begin{array}[t]{l}
\G_{ab}^{M N} M_{MN} \\[2mm] + 
\G_{ab}^{M_1 \ldots M_6} Z_{M_1 \ldots M_6}\ . 
\end{array}
\label{twelvealg}
\eea
The type IIB algebra in d=(1,9) reads

\phantom{.}\vspace{-10mm}
\bea 
\{ Q_{\a i} , Q_{\b j}\} &=&
\begin{array}[t]{l}
\G_{\a\b}^\m \Sigma^J_{ij}  Z_{J\m} \\[2mm]
{}+ \G_{\a\b}^{\m_1 \m_2 \m_3} \e_{ij} Z_{\m_1 \m_2 \m_3} \nn \\[2mm]
{}+ \G^{\m_1 \ldots \m_5}_{\a\b} \Sigma^J_{ij} Z_{J\m_1 \ldots \m_5} \ , \nn
\end{array}
\label{twob}
\eea
where we use the 
conventions $\S^J_{(ij)} = \e_{il} \S^{Jl}{}_j$, 
$\S_0 = -i \s^2, \S_1 = -\s^1$ and $\S_2 = \s^3$. 
These matrices satisfiy
$\S_I \S_J = 
\eta_{IJ} + 
\e_{IJ}{}^K \S_K = \eta_{IJ} +\e_{IJL}\eta^{LK} \S_K$,
with $\eta_{IJ} = (-++)$, $\e_{012}=1$,  
and hence generate $SL(2,{\bf R})$. 
In all cases the $Z$-charges fit the respective brane-scan, and
all cases form some decomposition of the $Q$-$Q$ part of 
$OSp(1|32)$.
\begin{table}[htb]
\setlength{\tabcolsep}{1mm}
\renewcommand{\arraystretch}{1.5}
\begin{center}
\begin{tabular}{|c||c|c|c|c|c|c|c|}
\hline 
D=11{} &{} &{}& 2 & {} & {} & 5 & {} {} {}\\[1mm]
\hline
IIA {}& $0_D$ & $1_F$  &  $2_D$ & {} &  $4_D$ & $5_S$ & $6_D$\\[1mm] 
\hline
IIB&{}{} &$1_F$, $1_D$&{} {} & \ \ $3_D^+$\ \ & {}{} & $5_S$, $5_D$ & {}{} \\[1mm] 
\hline
Type I {}&{} & $1_D$ & {} & {} & {} &  $5_D$  & {} {} {}\\[1mm] 
\hline
Het {}&{} & $1_F$ &{} &{} & {} & $5_S$ & {} {} {}\\[1mm] 
\hline 
\end{tabular}
\end{center}
{\bf Branescan}: the subscripts $F$, $D$, or
$S$,  denote `Fundamental', `Dirichlet' or `Solitonic' branes.
\end{table}

We will conjecture, for the purposes of this paper, that 
the rest of
the algebra completes (possibly some contraction of) $Osp(1|32)$:
\be
\begin{array}[t]{l}
\{ Q_{a} , Q_{b}\} = J_{ab} \nn \\[2mm] 
[ J_{ab},  Q_{c}] = - C_{c(a} Q_{b)} \nn \\[2mm] 
[ J_{ab},  J_{cd}] = 2 C_{(a(c} J_{b)d)\ ,}
\end{array} 
\label{ospalg}
\ee
which may be decomposed in terms of $SO(2,10)$ covariant
tensors to yield the extended
(1,9)-d superconformal algebra of 
van Holten and van Proeyen~\cite{vanPr}, namely
\be
\begin{array}[t]{l}
\{ Q_{a} , Q_{b}\} =
 - \frac{1}{128} \G_{ab}^{M N} J_{MN}  \nn \\[2mm]
\phantom{\{ Q_{a} , Q_{b}\} = } - \frac{1}{128 \cdot 6!}  
\G_{ab}^{M_1 \ldots M_6} J_{M_1 \ldots M_6} \\[2mm]
 [J_{MN} , Q_{a}] = - (\G_{MN})_{a}{}^{b} Q_b \nn \\[2mm]
 [J_{M_1 \ldots M_6} , Q_{a}] = 
 - (\G_{M_1 \ldots M_6})_{a}{}^{b} Q_b \nn \\[2mm]
 [J_{MN} , J^{KL}] = 8\ \de^{[K}_{[N} \ J_{M]}{}^{L]} \nn \\[2mm]
 [J_{MN} , J^{M_1 \ldots M_6}] = 
24\ \de^{[M_1}_{[N} \ J_{M]}^{\ \ M_2 \ldots M_6]} 
\label{twelvecomplete}
\end{array}
\ee
\bea
\begin{array}[t]{l}
 [J_{N_1 \ldots N_6} , J^{M_1 \ldots M_6}] =  \nn \\[2mm]
\phantom{ [J_{N_1 N_6}]} 
\begin{array}[t]{l}
{}-12 \cdot 6!\  \de^{[M_1 \ldots M_5}_{[N_1 \ldots N_5} \ J_{N_6]}^{\ \ M_6]} 
 \\[2mm]
{}+ 12 \ \e^{}_{N_1 \ldots N_6}{}^{[M_1 \ldots M_5|R|} \ J^{}_R{}^{M_6]} \ .
\end{array}
\end{array} \nn
\eea
In the (1,10)-d context this algebra was studied by D'Auria and
Fr\'{e}~\cite{Fre}.  We will try to take some first steps towards
constructing a conformal supergravity theory based on that type of algebra.

The signature of the vector space that appears in the above algebra is
$(2,10)$. This provides another hint of a connection to
string-theoretic ideas, as Vafa's~\cite{vafaevi} argument shows:
$Sl(2,Z)$-duality of type IIB strings may be explained via D-strings.
The zero-modes of the open strings stretched between such D-strings
determine the worldsheet fields of the latter. We have 
\bea
\Psi^\m_{-1/2} |k> & \m=0,1 & \hbox{2-d vect. field}\\
\Psi^m_{-1/2} |k> & m=2,\cdots,9 & \hbox{transv. scalars}
\label{openmodes}
\eea 
and hence we find on the D-string an extra U(1) gauge field. In
d=2 this is nondynamical, of course, but it leaves, 
after gauge fixing,
a pair of ghosts $B,C$ with central charge $c=-2$. The critical
dimension is hence raised by two, and the no-ghost
theorem~\cite{noghost,nogbrst}, which states that the BRST
cohomology effectively eliminates those extra dimensions, 
forces us to
assume the existence of a nullvector in the extra 
dimensions, and that
means they must have signature (1,1).

Taking the idea of strings moving in a 12-dimensional
target space more seriously, we are immediately led to the puzzle
of why strings oscillate in only 10 of these dimensions, but
never in the extra 2. If one has conformal symmetry in mind, 
there
is a natural answer: the 12 dimensions are those in which the
conformal group is linearly realized, but only a 10-dimensional
null hypersurface in real projective classes of these 
coordinates is 
physical. The extra two dimensions ``don't really exist''. 
The idea that strings might have some sort of target space 
conformal symmetry is not new~\cite{confstring}, but as of now no model
exists that can be convincingly linked to the string theories known 
today. 

At least part of the problem is the fact that a conventional
superconformal algebra in d=(1,9) does not exist, and while one can
write a conformal supergravity action, the fields one uses are subject
to differential constraints~\cite{tendimscg}. In contrast, for
d=(1,3), conformal N=1 supergravity and its superalgebraic
$SU(2,2|1)$-underpinnings are understood, and therefore we will
restrict ourselves to an analysis of $Osp(1|8)$, which may be
interpreted as a variant superconformal algebra. We note that
$SU(2,2|1)$ is not a subalgebra of $Osp(1|8)$. This is most clearly
seen by analyzing their embedding in $Osp(2|8)$~\cite{Fradlin}: let
the oscillators $a_{A} = (a^{K}, \ol a_{K}, a, \ol a)$ have the
(anti)commutation relations $[a^{K} ,\ol a_{L} ] = \de^{K}_{L}$,
$\{a,\ol a\}= 1$. Here $\ol a_{K} = \eta_{K \dot L} a^{\dot L}$ is up
to the $SU(2,2)$-metric $\eta_{K \dot L}$ the complex conjugate of
$a_{K}$. A real $Sp(8)$-spinor is represented by the complex pair
$(a^{K}, \ol a_{K}) = a_a$.  $Osp(2|8)$ has a total of 16 real
supersymmetry charges, namely the two $Sp(8)$-multiplets $Q^{(+)}_a =
a_a a = ( a^{K} a , \ol a_{K} a)$ and $Q^{(-)}_a = a_a \ol a = (
a^{K}\ol a , \ol a_{K}\ol a)$. The subalgebra $SU(2,2|1)$ is obtained
by selecting the supercharges $Q^{K} = a^{K} \ol a$ and $\ol Q_{K} =
\ol a_{K} a$, while $Osp(1|8)$ contains eight different supercharges,
namely $Q_a = a_a (a+\ol a)/\sqrt{2}$.  As a consequence, we obtain
the nonzero anticommutator $\{Q_K,Q_L\}=J_{KL}$. One might think
that we simply have to set $J_{KL} = 0 = J^{KL}$ in order to obtain 
the ordinary superconformal algebra. This true up to a factor -3 in 
the $Q^K$-$Q_L$ anticommutator, which is the trace of the 
$SU(2,2|1)$-metric.
In the oscillator representation this factor appears as
follows: the bosonic generators of $SU(2,2|1)$ are given by $J^K{}_L =
\half \{ a^K , \ol a_L \} - \frac{1}{8} \de^K_L \{a^N , \ol a_N\}$ and
$J = \half \{ a^K ,\ol a_K \} = \half [ a, \ol a]$ (which implies a
nontrivial trace condition on the total Hilbert space) and hence \be
\begin{array}[t]{rl}
\{Q^K , \ol Q_L\} 
\ = \ & \half \{ a^K ,\ol a_L \}  -  \half \de^K_L [ a, \ol a] \\[2mm]
\ = \ & J^K{}_L - \frac{3}{4}\de^K_L J \ ,
\end{array}
\ee
while for $Osp(1|8)$ we obtain
\be
\begin{array}[t]{rl}
\{Q^K, \ol Q_L\} \ = \ & \half \{ a^K , \ol a_L \} \\[2mm]
\ = \ & J^K{}_L + \frac{1}{4} \de^K_L J \ ,
\end{array}
\ee
where we have defined $J = \half \{ a^K ,\ol a_K \}$ in the 
same fashion. Apart from this factor, and of course the generators 
$J^{KL}= a^{(K} a^{L)}$ and $\ol J_{KL} = \ol a_{(K} 
\ol a_{L)}$, the two algebras are identical.

Let us present $Osp(1|8)$ in a $SO(2,4)$--covariant form.
First, we define $[a_{a}, a_{b}] = - C_{ab}$ with
$a,b=1,...,8$. We obtain $\{ Q_{a} , Q_{b}\} = a_{(a} a_{b)}$ 
and use the Fierz identity
\bea
 \de^c_{(a}  \de^d_{b)} & =
-\frac{1}{8} & \!\!\!\! \Big\{  \G^7{}_{ab}\ \G^{7cd}
+ \frac{1}{2} \G^{MN}_{ab}\  \G_{MN}^{cd} \nn \\ &&
{}+ \frac{1}{6} \G^{LMN}_{ab}\ \G_{LMN}{}^{cd} \Big\}
\eea
to rewrite this as
\bea
\{ Q_a , Q_b \} &=&
\begin{array}[t]{l}
\frac{1}{8} \Big\{  \G^7_{ab}\ a^c \G^7_c{}^d a_d \\
{}\ \ + \frac{1}{2} \G^{MN}_{ab}\ a^c \G_{MN}{}_c{}^d a_d  \phantom{\Big\}} \\
{}\ \ + \frac{1}{6} \G^{LMN}_{ab}\ a^c \G_{LMN}{}_c{}^d a_d \Big\}
\end{array}\\
 &\equiv&
\begin{array}[t]{r}
\frac{1}{4} \Big\{  \G^7_{ab}\ J_7  
+ \frac{1}{2} \G^{MN}_{ab}\ J_{MN} \\
{}+ \frac{1}{6} \G^{LMN}_{ab}\ J_{LMN} \Big\} \ .
\label{qq6}
\end{array}
\eea
The $SO(1,3)$-decomposition of the Gamma - matrices we use reads
\bea
\begin{array}{l}
\G^m=-\g^m\otimes\s^3,\G^7 = -\g^5\otimes \s^3 \\
\G^\oplus=\frac{1}{\sqrt{2}}\ \one \otimes\s^+,
\G^\ominus=\frac{1}{\sqrt{2}}\ \one \otimes\s^- \\
\{\Gamma^M,\Gamma^N\}=2\eta^{MN}= (--++++) , 
\end{array}
\label{rep6}
\eea
where $\G^M = \G^M{}_a{}^b$ 
and we have chosen $\eta_{\oplus \ominus} = 1$.
We raise and lower indices as
follows: $a^a = C^{ab} a_b$, 
$\G^\ast{}^{ab} = \G^\ast{}^a{}_c C^{cb} 
= C^{ac} \G^\ast{}_c{}^b $, 
$\G^\ast{}_{ab} = \G^\ast{}_a{}^c C_{cb} = C_{ac} 
\G^\ast{}^c{}_b $,
$C^{ac} C_{cb} = \de^a_b$.  
$C^{ab} = ( \g^0 \otimes \sigma^1 )^{ab}$ is the $8\times8$
charge conjugation matrix introduced above. 
With these conventions, among the matrices 
$\G^\ast{}_{ab}$ we find
$\G^7,\G^{MN}$ and $\G^{MNP}$ symmetric
under interchange of $a$ and $b$, while
$C,\G^M,\G^{MNPQ}$ and $\G^{MNPQR}$ are antisymmetric.
Similarly, the real $4\times4$ matrices 
$\g^\ast{}_{\a\b}$ are split into
the symmetric $\g^m,\g^{mn}$ and the antisymmetric 
$C^4,\g^{mnp}$ and $\g^5$.
The remaining sectors of $Osp(1|8)$ 
\bea
[ J^\ast , Q_a ] &=&
\begin{array}[t]{r}
-\G^\ast{}_a{}^b Q_b 
\end{array}\label{quirk}\\[0mm]
 [ J^7 , J^{MNP} ] &=&
\begin{array}[t]{r}
\frac{1}{3} \epsilon^{MNPRST} J_{RST}
\end{array}\\[0mm]
 [ J^{MN} , J_{RS} ] &=&
\begin{array}[t]{r}
8\ \de^{[N}_{[R} J^{M]}{}_{S]}
\end{array}\\[0mm]
 [ J^{MN} , J_{RST} ] &=&
\begin{array}[t]{r}
12\ \de^{[N}_{[R} J^{M]}{}_{ST]}
\end{array}\\[0mm]
 [ J^{MNP} , J_{RST} ] &=&
\begin{array}[t]{l}
2\ \epsilon^{MNP}{}_{RST} J^7\\{} - 36\ 
\de^{[M}_{[R}\de^{N}_{S} J^{P]}{}_{T]}
\end{array}\label{quark}
\eea
are now straightforward to decompose under $SO(1,3)$,
and we use then the notation
\newpage

\be
\begin{array}{ll}
P^m=J^{\oplus m}&
K^m=J^{\ominus m}\\
E^{mn}=J^{\oplus mn}&
F^{mn}=J^{\ominus mn}\\
M^{mn}=\frac{1}{2}J^{mn} &
D = J^{\oplus\ominus} \qquad \quad  A = J^7\\
V^m = J^{\oplus \ominus m}&Z^m = -\frac{1}{3!} \e^{mnpq} J_{npq} \ .
\end{array}
\label{decomp}
\ee

\section{Curvatures}

In $SO(2,4)$-covariant language
the connection 1-forms are written as 
$h = h_7 J^7 + \half h_{MN} J^{MN}
+ \frac{1}{3!}  h_{MNP} J^{MNP} + \psi^a Q_a$, with 
$\psi^a = (\phi^\a , \psi^\a)$,
and the curvatures $R = dh + hh$
are given by
\setlength{\arraycolsep}{0mm}
\be
R = 
\begin{array}[t]{l} 
\begin{array}[t]{l}
\Big\{ dh_7 + \frac{1}{8} \psi^a \G_{ab} \psi^b \\
{}+\frac{1}{36} \e^{MNPRST} h_{MNP}h_{RST} \Big\} J^7
\end{array}
\\ 
{}+\half 
\begin{array}[t]{rl}
\Big\{ dh_{MN}  + 2 h_{MK}h^K{}_{N}&\\ 
{}- h^{RS}{}_M h_{RSN}&\\
{}+ \frac{1}{8} \psi^a \G_{MNab} \psi^b &\Big\} J^{MN}
\end{array}
\\ 
{}+\frac{1}{6}
\begin{array}[t]{rl}
\Big\{ dh_{MNP} + 6 h_{M}{}^K h_{KNP}&\\
{}+ \frac{1}{3} \e_{MNPRST} h^{RST} h_7 &\\
+ \frac{1}{8} \psi^a \G_{MNPab} \psi^b &\Big\} J^{MNP}
\end{array}
\\ 
{}+
\begin{array}[t]{l}
\Big\{ d\psi^a + h_7 \G^{7a}{}_b \psi^b \\ {}+
\frac{1}{2} h_{MN} \G^{MNa}{}_b \psi^b\\ {}+  
\frac{1}{6} h_{MNP} \G^{MNPa}{}_b \psi^b \Big\} Q_a \ .
\end{array}
\end{array}
\label{sixdecomp}
\ee
The gauge transformations $\de h = d\l + [h,\l]$
imply $\de R = [R,\l]$, i.e.
\bea
\de R\  &=&
\begin{array}[t]{rl}
&
\begin{array}[t]{r}
\Big\{ \frac{1}{18} \e^{MNPRST} R_{MNP}\l_{RST} \\ 
{}- \frac{1}{4} R^a \G^7{}_{ab} \l^b \Big\} J^7
\end{array}
\\ 
+\half 
& \begin{array}[t]{r}
\Big\{ 4 R_{MK}\l^K{}_{N} - 2 R^{RS}{}_M \l_{RSN}\\
{}- \frac{1}{4} R^a \G_{MNab}\l^b \Big\} J^{MN} 
\end{array} 
\\ 
{}+\frac{1}{6}
& 
\begin{array}[t]{rl}
\Big\{ - \frac{1}{3} \e_{MNPRST} R_7 \l^{RST} & \\ 
{}+ \frac{1}{3} \e_{MNPRST} R^{RST} \l_7 & \phantom{\Big\}} \\ 
{}+ 6 \R_{M}{}^K \l_{KNP} &\phantom{\Big\}}\\
{} - 6 \R_{MN}{}^K \l_P{}^K &\phantom{\Big\}} \\
{}- \frac{1}{4} R^a \G_{MNPab} \l^b & \Big\} J^{MNP} 
\end{array} \\
+&
\begin{array}[t]{rl}
\Big\{ R_7 \G^{7a}{}_b \l^b  + \frac{1}{2} R_{MN} \G^{MNa}{}_b \l^b  & \\
{} + \l_7 \G^{7a}{}_b R^b + \frac{1}{2} \l_{MN} \G^{MNa}{}_b R^b &
\phantom{\Big\}}  \\
{} + \frac{1}{6} R_{MNP} \G^{MNPa}{}_b \l^b &\phantom{\Big\}} \\
{}+ \frac{1}{6} \l_{MNP} \G^{MNPa}{}_b R^b & \Big\} Q_a\ .
\end{array}
\label{sixdgaugetrans}
\end{array}
\eea
The $SO(1,3)$-decomposition results in
\bea
R(P)^m &\ =\ &
\begin{array}[t]{l}
de^m+\o^m{}_ne^n+2be^m-2E^m{}_nv^n \\
{}-2\wt{E}^{m}{}_nz^n-\frac{1}{4\sqrt{2}}\psibar\g^m\psi
\end{array}
\label{curve}\\
R(E)^{mn}&=&
\begin{array}[t]{l}
            dE^{mn}-2\o^{[m}{}_kE^{n]k}+2bE^{mn}\\                
            +2\wt{E}^{mn}a-4e^{[m}v^{n]}
            +2\e^{mnpq}e_pz_q\\
            +\frac{1}{4\sqrt{2}}\psibar\g^{mn}\psi
\end{array}\\
R(Q)&=&
\begin{array}[t]{l}
d\psi+\big(-a\g^5+b+\frac{1}{4}\o^{mn}\g_{mn}\\ 
         -v^m\g_m+z^m\g^5\g_m\big)\psi\\
      +\left(\sqrt{2}e^m\g_m+\frac{1}{\sqrt{2}}E^{mn}\g_{mn}
      \right)\phi
\end{array}\label{ospqcurv}\\
R(M)^{mn}&=&
\begin{array}[t]{l}
d\o^{mn}-\o^{[m}{}_k\o^{n]k}+\\4v^{[m}v^{n]}+4z^{[m}z^{n]}
           \\  -8e^{[m}f^{n]}
          -8E^{[m}{}_kF^{n]k}\\+\frac{1}{2}\psibar\g^{mn}\phi
\end{array}\\
R(D)&=&
\begin{array}[t]{l}
db-2e^mf_m-E^{mn}F_{mn}\\ +\frac{1}{4}\psibar\phi
\end{array}\\
R(A)&=&
\begin{array}[t]{l}
da-2v^mz_m-\wt{E}^{mn}F_{mn}\\
       +\frac{1}{4}\psibar \g^5\phi
\end{array}\\
R(V)^m&=&
\begin{array}[t]{l}
dv^m+\o^m{}_nv^n+2z^ma\\+2E^m{}_nf^n-2F^m{}_ne^n\\
         +\frac{1}{4}\psibar\g^m\phi
\end{array}\\
R(Z)^m&=&
\begin{array}[t]{l}
dz^m+\o^m{}_nz^n-2v^ma\\+2\wt{E}^{mn}f_n
         +2\wt{F}^{mn}e_n\\+\frac{1}{4}\psibar \g^5\g^m\phi
\end{array}\\ 
R(S)&=&
\begin{array}[t]{l}
d\phi+\big(a\g^5-b+\frac{1}{4}\o^{mn}\g_{mn}\\-v^m\g_m
          -z^m\g^5\g_m\big)\phi\\
         +\big(-\sqrt{2}f^m\g_m+
\frac{1}{\sqrt{2}}F^{mn}\g_{mn}\big)\psi
\end{array}\label{ospscurv}\\
R(K)^m&=&
\begin{array}[t]{l}
df^m+\o^m{}_nf^n-2bf^m\\+2F^m{}_nv^n-2\wt{F}^{mn}z_n\\
         +\frac{1}{4\sqrt{2}}\phibar\g^m\phi
\end{array}\\            
R(F)^{mn}&=&
\begin{array}[t]{l}
dF^{mn}-2\o^{[m}{}_kF^{n]k}-2bF^{mn}\\                
            -2\wt{F}^{mn}a+4f^{[m}v^{n]}\\
         +2\e^{mnpq}f_pz_q
            +\frac{1}{4\sqrt{2}}\phibar\g^{mn}\phi\ .
\end{array}
\label{curvF}
\eea
The duals\footnote{We remind the reader that in Minkowski space
.} 
are defined by 
$\widetilde{X}^{mn}\equiv(1/2)\e^{mnpq}X_{pq}$, 
$\wt{\wt{X}}^{{}_{{}_{{}_{{}_{{}_{\scriptstyle mn}}}}}}=-X^{[mn]}$
and the bars on the Majorana fermions are defined by $\psibar=\psi^\top
C_4=\psi^\top\g^0$.

In order to obtain the curvatures of
a $SU(2,2|1)$-gauge theory, we simply drop 
the fields $E_{mn}$, $v_m$, $z_m$ and $F_{mn}$ and
change the fermionic curvatures to

\phantom{.}\vspace{-10mm}
\bea
R(Q)&=&
d\psi+(3a\g^5+b+\frac{1}{4}\o^{mn}\g_{mn})\psi \nn \\
     &&     +\sqrt{2}e^m\g_m\phi  \\
R(S)&=&
d\phi+(-3a\g^5-b+\frac{1}{4}\o^{mn}\g_{mn})\phi \nn \\
 && -\sqrt{2}f^m\g_m\psi \ .
\eea
The only difference to\rf{ospqcurv} and\rf{ospscurv} is the additional 
factor $-3$ in the terms $a\g^5\psi$ and $a\g^5\phi$.

\section{Actions}

We now construct an affine action quadratic in curvatures~\cite{MacDowell} 
and invariant under the symmetries $S$, $K_m$ and $F_{mn}$. By affine we
mean that no vierbeins are used to contract indices, but only constant
Lorentz tensors such as $\e^{\m\n\r\s}$, $\eta_{\m\n}$ and Dirac
matrices. The most general parity-even, Lorentz-invariant,
dilaton-weight zero, mass dimension zero affine action
($S=\int_{M}\cal{L}$ for some four--manifold $M$) reads
\setlength{\arraycolsep}{1mm}
\bea
-\cal{L}&=&\a_0\e_{mnpq}R(M)^{mn}R(M)^{pq}\nn\\ 
         &&{}+\a_1R(A)R(D)+\a_2R(V)^mR(Z)_m\nn\\
         &&{}+\a_3\e_{mnpq}R(E)^{mn}R(F)^{pq}\nn\\
         &&{}+\b\widebar{R(Q)}\g^5R(S)\ .
\label{action}
\eea
This action
\footnote{
The action is Hermitean and the curvatures are
real if one takes the reality condition for Majorana spinors
$\overline{\psi}=\psi^\top C_4=\psi^\dagger i\g^0$. We denote the
left hand side of the Minkowski action in\rf{action} by $-\calL$
to stress that we are using the metric $(-+++)$ rather than the
Euclidean notation of~\cite{Kaku1}. The sign $-\calL$ ensures
that the kinetic terms for the vierbein have the correct sign,
see, for example, reference~\cite{Fradkin}.
}
is, of course, manifestly general coordinate invariant since the
integration measure $\e^{\m\n\r\s}$ is a tensor density under general
coordinate transformations. The term $\a_2 R(V)^m R(Z)_m$ is not
$A$-invariant like all the other terms and one could therefore
consider setting the coefficient $\a_2=0$ already at this point.
Since we are interested in a theory of gravity we set $\a_0=1$ (in
fact, no nontrivial solution exists for $\a_0=0$).

The requirement that the action in\rf{action} be invariant
under the symmetries $S,K$ and $F$ yields
\be
\a_0=1,\a_1=-32,\a_2=0,\a_3=8,\b=-8,\label{coeffs}
\ee
as well as the following constraints on the field strengths:

\phantom{.}\vspace{-10mm}
\bea
R(P)^m&=&0\label{c1}\\
R(E)^{mn}&=&-{}_*\widetilde{R}(E)^{mn}\label{c2}\\
R(Z)^m&=&{}_*R(V)^m\label{c3}\\
R(Q)&=&-\g^5{}_*R(Q)\ , \label{c4}
\eea
which are in turn invariant under $S$, $K$ and $F$. The sign on the right
hand side is in principle at our disposal. The above choice guarantees
that the constraints can be solved algebraically if the vierbein is assumed
to be invertible. 

In order to compare with the $SU(2,2|1)$-case we
again write down the most general parity even,
dilaton weight zero, affine action
and fix the coefficients and constraints by
requiring invariance with respect to the symmetries 
$K$ and $S$. The results 
are
\bea
-{\cal L}&=&\e_{mnpq}R(M)^{mn}R(M)^{pq}\nn \\ &&{}+32 R(A)R(D)
         -8\overline{R(Q)}\g^5R(S)
\label{sS}
\eea
with the constraints
\setlength{\arraycolsep}{1mm}
\bea
R(P)^m & = &0 \label{sc0}\\
R(Q)&=&-\g^5{}_*R(Q)\label{sc1}\\
R(A)&=&{}_*R(D) \ .\label{sc2}
\eea 
These constraints are again invariant under the symmetries $S$ and $K$ and
algebraically solvable. We note that the actions\rf{sS} and\rf{action}
are the most complicated ones in a series of gauge theories covering 
Anti-de-Sitter gravity based on $Sp(4)$, its supersymmetrized 
$Osp(1|4)$-version and of course ordinary conformal gravity based 
on $SU(2,2)$. The complexity of the constraints increases with the size 
of the algebra, however in each case, a kinematical study of the gauge
algebra shows that the constraints are exactly such that gauge 
transformations $\delta h=d\lambda+[h,\lambda]$ are modified precisely so
that the gauge algebra closes onto general coordinate transformations, rather
than $P^m$ gauge transformations
(i.e. gauge transformations generated by the
translation generator $P_m$)~\cite{pvn}. In fact one can adopt a purely
kinematical approach in which one derives the constraints through the
requirement that the algebra closes onto general coordinate transformations.
If one makes a similar kinematical study of the $Osp(1|8)$ algebra, one is
quickly led to the conclusion that no set of algebraically solvable
constraints exists such that the algebra closes onto general coordinate
transformations. However, in the hope that the model could again be made
consistent through further generalizations of the $E_{mn}$ gauge symmetries
and super gauge symmetries ($Q$) along with the usual trade between $P_m$
gauge transformations and general coordinate transformations, we followed the
dynamical affine action approach which has enjoyed considerable success as
evidenced by the string of models given above.

\section{Constraints}

The constraints\rf{sc0} --\rf{sc2} are necessary but not yet sufficient
for obtaining conformal supergravity. For an irreducible representation
of the conformal superalgebra we should try and express as many fields as 
possible algebraically in terms of a minimal set. In the conformal case,
the {\sl maximal} set of solvable constraints is
\bea
0&=&R(P)_{\m\n}{}^m\\
0&=&\wt{R}(M)^{mn}e_n+2R(A)e^m
\nn\\&&{}-\frac{1}{2\sqrt{2}}
\psibar \g^5\g^mR(Q),
\label{sfeqn}
\eea
as well as 
\be
\g^\m R(Q)_{\m\n}=0 \ , \label{sc4}
\ee
which can be shown~\cite{Kaku1} to be necessary for
$Q$-supersymmetry of the action\rf{sS}.

For the $Osp(1|8)$-case, we summarize in figure~\ref{constraints}
the curvature components that, when constrained to zero, lead to 
algebraic equations for connection pieces. 
\begin{figure}[ht]
\setlength{\tabcolsep}{0.5mm}
\begin{tabular}{|ccccccccccccccc|}
\hline
&&&&&&&&&&&&&&\\[-4mm]
$R(P)_{\m\n}{}^m $&=&\ul{24}&=&\ul{16}&+&$\wt{\ul{4}}$&+&\ul{4}&&&&&&\\
&&&&$\surd$&&$\surd$&&$\surd$&&&&&&\\[1mm]
\hline
&&&&&&&&&&&&&&\\[-4mm]
$R(E)_{\m\n}{}^{mn}$&=&
\ul{36}&=&$\wt{\ul{1}}$&+&\ul{10}&+&$\wt{\ul{9}}$&+&\ul{9}&+&\ul{6}&+&\ul{1}\\
&&&&$\surd$&&$\times$&&$\surd$&&$\surd$&&$\surd$&&$\surd$\\[1mm]
\hline
&&&&&&&&&&&&&&\\[-4mm]
$R(Q)_{\m\n}$&=&
\ul{24}&=&\ul{8}&+&\ul{12}&+&\ul{4}&&&&&&\\
&&&&$\times$&&$\surd$&&$\surd$&&&&&&\\[1mm]
\hline
&&&&&&&&&&&&&&\\[-4mm]
$R(M)_{\m\n}{}^{mn}$&=&
\ul{36}&=&$\wt{\ul{1}}$&+&\ul{10}&+&$\wt{\ul{9}}$&+&\ul{9}&+&\ul{6}&+&\ul{1}\\
&&&&$\times$&&$\times$&&$\times$&&$\surd$&&$\surd$&&$\surd$\\[1mm]
\hline
&&&&&&&&&&&&&&\\[-4mm]
$R(D)_{\m\n}$&=&\ul{6}&&&&&&&&&&&&\\
&&$\surd$&&&&&&&&&&&&\\[1mm]
\hline
&&&&&&&&&&&&&&\\[-4mm]
\multicolumn{7}{|c}{$R(V)_{\m\n}{}^m + {}_*R(Z)_{\m\n}{}^{m}$}&=&
\ul{24}&=&\ul{16}&+&$\wt{\ul{4}}$&+&\ul{4}
\\
\multicolumn{7}{|c}{}&&&&$\surd$&&$\surd$&&$\surd$
\\[1mm]
\hline
&&&&&&&&&&&&&&\\[-4mm]
\multicolumn{7}{|c}{$R(V)_{\m\n}{}^m - {}_*R(Z)_{\m\n}{}^{m}$}&=&
\ul{24}&=&\ul{16}&+&$\wt{\ul{4}}$&+&\ul{4}
\\
\multicolumn{7}{|c}{}&&&&$\times$&&$\surd$&&$\surd$
\\[1mm]
\hline
\end{tabular}
\vspace{-8mm}
\caption{
Lorentz irreducible pieces of the ``solvable''
curvatures. The ticks ``$\surd$'' and crosses ``$\times$''
indicate those Lorentz irreducible pieces of
curvatures that may or may not,
respectively, be solvably constrained.\label{constraints}} 
\end{figure}
We have found the following maximal set of
solvable constraints:
\bea
0&=&R(P)_{\m\n}{}^m\label{mc1}\\
0&=&R(E)_{\r[\m \n]}{}^\r\label{mc2}\\
0&=&R(E)_{\m\n}{}^{\n\m}\\
0&=&\e^{\m\n\r\s}R(E)_{\m\n\r\s}\\
0&=&R(E)_{\m\n}{}^{mn}+{}_*\wt{R}(E)_{\m\n}{}^{mn}\label{mc25}\\
0&=&R(Z)_{\m\n}{}^m-{}_*R(V)_{\m\n}{}^m\label{mc3}\\
0&=&\g^\m R(Q)_{\m\n}\label{mc4}\\
0 &=& R(M)_{\r\m\n}{}^\r-\frac{1}{2}g_{\m\n}R(M)_{\r\s}{}^{\s\r}
\nn\\&&{}+2{}_*R(A)_{\m\n}+\frac{1}{2\sqrt{2}}\overline{R(Q)}_{\r\n}
\g_\m\psi^\r 
\nn\\&&{}-2{}_*R(E)_{\r\n\s\m}z^{\r\s}
-2R(E)_{\r\n\s\m}v^{\r\s}\nn\\&&{}+2R(V)_{\r\n\s}E^{\r\s}{}_\m
+2R(Z)_{\r\n\s}\wt{E}^{\r\s}{}_\m\ . \label{mc5}
\eea
All further constraints follow from this set, either
algebraically or, for example, by Bianchi identities.
Unlike the superconformal case, this set does not guarantee
all the symmetries necessary for consistency of the action\rf{action}. 

\section{Problems}

The set\rf{mc1} --\rf{mc5} of constraints does not allow us to express
explicitly all fields in terms of a minimal set. Rather, we obtain a
coupled set of equations, which determine, say, $\o_\m{}^{mn}$,
$F_\m{}^{mn}$, $\phi_\m^\a$, $z_{(\m\n)}$, $v_{(\m\n)}$ and 
$z_{[\m\n]}+{}_*v_{[\m\n]}$ in terms of $e_\m{}^m$, $E_\m{}^{mn}$, $\psi_\m$,
$b_\m$, $a_\m$ and $z_{[\m\n]}-{}_*v_{[\m\n]}$. We may try to solve
these equations iteratively, but in order to investigate the
symmetries of the action, an explicit solution is not necessary.
However, then the same problem occurs when we examine the
invariance of the action under the remaining symmetries $V,Z,E$ and
$Q$. These symmetries need to be modified when acting on dependent
fields such that they leave\rf{mc1} --\rf{mc5} invariant, and again
the constraints provide only a coupled set of equations for the extra
transformations of the dependent fields. In order to calculate
further, one can make a consistent expansion in the number of fields
and study the model in the lowest order in this expansion.

Let us consider the constraint $R(P)^m=0$ and some $\de\in \{V,Z,E,Q\}$. 
On independent fields $\de$ acts simply as a gauge transformation
\be
\de h^A_{\rm Indept.}=d \e^A+\e^C h^Bf_{BC}{}^A
=\de_{\rm Group}h^A_{\rm Indept.}. 
\ee
However acting on dependent fields we have
\bea
\de h^A_{\rm Dept.}&=& d\e^A+\e^C h^Bf_{BC}{}^A+\what{\de}h^A_{\rm Dept.}
\nn \\
&\equiv& \de_{\rm Group}h^A_{\rm Dept.}+\what{\de}h^A_{\rm Dept.}\ , 
\eea
where the extra transformations 
$\what{\de}h^A_{\rm Dept.}$ are determined by
requiring that the constraints are invariant under $\de$, for example
\bea
0=\de R(P)^m&=&\de_{\rm Group}R(P)^m+\what{\de}R(P)^m\nn \\
&=&\de_{\rm Group}R(P)^m+\dehat\o^{mn}e_n \nn \\
&&{}-2E^{mn}\dehat v_n-2\wt{E}^{mn}\dehat z_n\ .\label{solve}
\eea
Note that in\rf{solve}, the extra transformations of 
three dependent fields appear, so that neither is uniquely determined.
In contrast, for conformal supergravity only $\dehat \o^{mn}$ is present
and can then be determined.
One may write down similar expressions for all other
constraints which {\it in principle} 
uniquely determine all extra transformations of the 
dependent fields. 

In practice, to write down a solution for the 
extra transformations of the dependent fields, we solve the set
of coupled equations for the extra transformations iteratively.
Namely, we make an expansion order by order in the number of
independent fields, where one counts the vierbein as a
Kronecker delta (i.e. field number zero). This means
that to first order, we ignore  the terms 
$-2E^{mn}\dehat v_n-2\wt{E}^{mn}\dehat z_n$ in\rf{solve}.
To linear order one then obtains
\bea
e^n\dehat_{V}\wt{\o}_{mn}&=&-2\wt{R}(E)_{mn}\e^n\label{me}\\
e^n\dehat_{Z}\wt{\o}_{mn}&=&2R(E)_{mn}\e^n\\
e^m\g_m\dehat_{V}\phi&=&-\frac{1}{\sqrt{2}}\g_mR(Q)\e^m\\
e^m\g_m\dehat_{Z}\phi&=&+\frac{1}{\sqrt{2}}\g^5\g_mR(Q)\e^m
\label{you}
\eea
and our action is then indeed invariant under $V$ and $Z$.
However, at linear order, the action is not invariant under
$E_{mn}$ and $Q$ symmetries although many terms do cancel. The
$E$-transformations leading to non-vanishing variations are
\setlength{\arraycolsep}{0.5mm}
\bea
\dehat_{E}\phi_\m&=&-\frac{1}{4}[\g^\r\g^{mn}R(S)_{\m\r} \nn\\&&
                    {}+\frac{1}{6}\g_\m\g^{\s\r}\g^{mn}R(S)_{\r\s}]\e_{mn}\\
\dehat_{E}\o^0_{\m mn}&=&R(V)_{mn}{}^{r}\e_{r\m}-
                         R(V)_{\m m}{}^{r}\e_{rn}\nn\\&&{}+
                         R(V)_{\m n}{}^{r}\e_{rm}
                         +R(Z)_{mn}{}^{\r}{}_*\e_{\r\m}\nn\\&&{}-
                         R(Z)_{\m m}{}^{r}\wt{\e}_{rn}+
                         R(Z)_{\m n}{}^{r}\wt{\e}_{rm}\nn\\&&{}
                        -\frac{2}{3}e_{\m[m}\dehat\o_{n]}\\
\dehat_{E}\o_\m&=&2(1-\b)(\wt{\e}_{mn}R(Z)^n\nn\\&&
                  {}+\e_{mn}R(V)^{n})\\
\dehat_{E}v_{[\m\n]}&=&\frac{1}{8}(1-\a)[R(M)_{\r\s\m\n} \e^{\r\s}\nn\\&&
                  {}+8R(D)_{\r[\m}\e_{\n]}{}^\r]\\
\dehat_{E}z_{[\m\n]}&=&\frac{1}{8}\a [R(M)_{\r\s\m\n}{}_*\e^{\r\s}\nn\\&&
                  {}+8R(D)_{\r[\m}{}_*\e_{\n]}{}^\r] \\
\dehat_{E}v_{(\m\n)}&=&\frac{1}{6}g_{\m\n}R(D)_{\r\s}\e^{\r\s}\\
\dehat_{E}z_{(\m\n)}&=&\frac{1}{6}g_{\m\n}R(D)_{\r\s}{}_*\e^{\r\s},
\label{elvis}
\eea
while the relevant extra $Q$-transformations read
\bea
\dehat_Q\phi_\mu&=&
\frac{1}{3\surd 2}[R(D)_{\m\r}+R(A)_{\m\r}\g^5\nn\\&&{}
+\frac{1}{2}(R(V)_\r+R(Z)_\r\g^5)\g_\m]\g^\r \e \label{sun}\\
\dehat_Qv_{\m\n}&=&\frac{1}{8\surd
2}(1-\a)\overline{R(Q)}\e\\
\dehat_Qz_{\m\n}&=&\frac{1}{8\surd
2}\a\overline{R(Q)}\g^5\e \ .
\eea
This means the action\rf{action} is not consistent. In a flat
gravitational and otherwise trivial background the fields $\psi^\a$
and $E^{mn}$ enter the quadratic part of the action only in terms of
their linearized field strengths $d\psi^\a$ and $dE^{mn}$. Hence the
associated gauge invariances are necessary for obtaining invertible kinetic
terms. Since they do not survive at the interacting level, we conclude
that the theory does not exist in the way we have formulated it, unless one
can find generalizations of $Q$ and $E_{mn}$ symmetries under which the
action is invariant. 

\section{Conclusions}

Even though $Osp(1|8)$ seems to fit naturally into a pattern
of (super)gravity theories in d=(1,3), the affine action has
serious deficiencies. Since the affine action does not suffice either
for theories with a higher number of supersymmetries, we may speculate 
that one should add an appropriate number of non-gauge fields. 
This is also borne out by the spectrum of (conformal) supergravities 
in high dimensions. At this point is it not clear precisely what
we should add and how one systematically derives then appropriate
constraints. Work on these issues is in progess.  


\end{document}